\documentclass[conference]{IEEEtran}
\usepackage{indentfirst}
\usepackage{cite}
\usepackage{mathrsfs}
\usepackage{stfloats}
\usepackage[english]{babel}
\usepackage{blindtext}
\usepackage{algorithm}
\usepackage{algorithmic}
\usepackage{graphicx}
\usepackage{bm}
\usepackage{subfigure}
\usepackage{amsmath}
\usepackage{amssymb}
\usepackage{cases}
\usepackage{url}
\usepackage{setspace}
\usepackage{multirow} 

\allowdisplaybreaks[4]

\begin{document}
%
\title{Energy Efficient Resource Allocation for Control Data Separation Architecture based H-CRAN with Heterogeneous Fronthaul}

\author{\IEEEauthorblockN{Qiang Liu, Gang Wu, Yingchu Guo, and Yusong Zhang}
\IEEEauthorblockA{National Key Laboratory of Science and Technology on
 Communications\\
University of Electronic Science and Technology of China, Chengdu 611731, China\\
Email: liuqiang12040913@gmail.com, wugang99@uestc.edu.cn}}


\maketitle

\begin{abstract}
Control data separation architecture (CDSA) is a more efficient architecture to overcome the overhead issue than the conventional cellular networks, especially for the huge bursty traffic like Internet of Things, and  over-the-top (OTT) content service. In this paper, we study the optimization issue of network energy efficiency of the CDSA-based heterogeneous cloud radio access networks (H-CRAN) networks, which has heterogeneous fronthaul between control base station (CBS) and data base stations (DBSs). We first present a modified  power consumption model for the CDSA-based H-CRAN, and then formulate the optimization problem with constraint of overall capacity of wireless fronthaul. We work out the resource assignment and power allocation by the convex relaxation approach Using fractional programming method and Lagrangian dual decomposition method, we derive the close-form optimal solution and verify it by comprehensive system-level simulation. The simulation results show that our proposed algorithm has $8\%$  EE gain compared to the static algorithm, and the CDSA-based H-CRAN networks can achieve up to $16\%$ EE gain compared to the conventional network even under strict fronthaul capacity limit.
\end{abstract}


%

\section{Introduction}
The fifth generation mobile communication system (5G) is expected to support massive connections with either higher data rate, or lower latency, or ultra-higher reliability, or higher energy effiicency \cite{ref1} than the fourth generation mobile communication system (4G) like 3GPP long-term-evolution (LTE). To achieve those goals, many concepts, including small cell, heterogeneous network (HetNet), and cloud radio access network (Cloud-RAN) \cite{ref2}, have been proposed and studied. As indicated in  \cite{ref3}, \cite{ref4}, heterogeneous cloud RAN (H-CRAN), which combines cloud computing with HetNet, has been regarded as one promising network architecture to overcome the limitations of conventional HetNet and achieves high spectral efficiency (SE) and energy efficiency (EE).

Despite the promising advantages of H-CRAN, the most significant challenge facing its implementation is the limited capacity of the fronthaul link\cite{ref5}, which connect baseband units (BBUs) and remote radio heads (RRHs) via either wired or wireless medium. Although optical fiber is recognized as the ideal medium and capable of supporting huge capacity, toward ultra densier network deployment of 5G, it become difficult to build a fiber-link because of unpredictable expense. Meanwhile, with the development of ultra-high data rate millimeter-wave (mmWave)  wireless transmission techniques, wireless fronthaul solution is becoming an attractive deployment solution\cite{ref6}, which make the heterogeneous fronthaul \cite{ref7} a potential solution to realize H-CRAN.

The issue of limited backhaul has been discussed in \cite{ref8}, \cite{ref9}, and \cite{ref10}. In \cite{ref8}, Derrick \emph{et al.} investigated the downlink resource allocation problem to maximize EE with limited backhaul capacity for each BS. They formulated the optimization problem for the zero-forcing beamforming (ZFBF) transmission, and derived the closed-form optimal power allocation solution. But this work is limited to the conventional homogenenous multi-cell network, not for HetNet. In \cite{ref9}, Vu \emph{et al.} considered the joint coordinated beamforming and admission control issue in downlink C-RAN with fronthaul capacity limits, by minimizing transmission power of overall network. Using convex relaxation approach, they developed an iterative algorithm to solve the optimization problem. Dhifallah \emph{et al.} in \cite{ref10} studied the EE issue of  C-RAN with heterogeneous backhaul, and took the total  power consumption of network as the optimization goal. But \cite{ref10}  only investigated the impact of wireless and wired fronthaul with respect to the maximum data rate. 

In conventional cellular networks, its functionalities for ubiquitous access provision and data service provision are tightly coupled. With the intensive deployment of small cells, problems like serious signaling interference and huge signaling overhead, will challenge the implementation of 5G networks \cite{ref11}. As a potential solution, control data separation architecture (CDSA) is proposed to cope with those problems, which provides a flexible deployment solution \cite{ref12}. In the CDSA networks\cite{ref14},\cite{ref15}, control base station (CBS) provides pervasive coverage for all user equipment (UEs) and data base stations (DBSs) are used to support data rate transmission \cite{ref13}. With the separation of control plane and data plane, the overhead cost triggered by massive UE connection and mobility handover can be significantly decreased \cite{ref11}. Meanwhile, the CDSA networks enable network topology to accommodate the variation of traffic demand dynamically.

Energy efficiency of the CDSA networks is totally different from the conventional networks, which have not been widely investigated. In \cite{ref11}, Xu \emph{et al.} indicated that the CDSA networks can save up to one third of the power consumption in urban deployment scenarios considering variation of traffic demand by jointly designing the sleep strategy of DBSs and control overhead reduction. Furthermore, using stomachic geometry theory, the SE/EE tradeoff of the CDSA networks is theoretically studied by Zhang \emph{et al.} in  \cite{ref16}, where the reduction of control signals of DBSs and the increment of control signals of CBS were calculated independently. The results indicated that the CDSA networks effectively solve the overhead problem and could achieve higher SE and EE than th the conventional networks. It is worthy noting that \cite{ref16} still use the conventional power consumption model without considering feature of H-CRAN. 

Actually, signal processing and backhaul in H-CRAN play a significant role in reducing system power consumption. So far as we know, the energy efficient resource allocation problem of the CDSA networks have not been studied yet, especially with heterogeneous fronthaul. Different from those work mentioned above, in this paper, we focus on the network energy efficiency of downlink CDSA-based H-CRAN based on heterogeneous fronthaul. We first build a modified power consumption model for the CDSA-based H-CRAN networks. And we take the network energy efficiency  as optimization goal. To obtain the optimal resource assignment and power allocation, we formulate the optimization problem with the constraint of average minimum data rate requirement and capacity limit of wireless fronthaul, which is totally different from previous work. In order to solve the proposed non-convex problem, the objective function is transformed by fractional programming method. Using the Lagrangian dual decomposition method, we derive the close-form expression of optimal power allocation and resource assignment even the optimization problem is non-convex. As the simulation results show, our proposed algorithm achieves maximum $8\%$ EE gain compared to static algorithm, and that the CDSA networks have maximum $16\%$ EE gain compared to the conventional networks.

The reminder of this paper is organized as follows. In Section II, we describe the system model of CDSA-based H-CRAN, including the network model and the energy consumption model. Furthermore, the optimization problem is described. In Section III, the energy efficient resource allocation algorithm is developed. The simulation results in Section IV verify the effectiveness of our proposed algorithm. Finally, we draw the conclusion in Section V.
\section{System Model And Problem Formulation}
\subsection{System Model}
\subsubsection{Network Model}
We consider a 2-tier downlink H-CRAN with CDSA architecture, consisting of one macro-cell control base station (CBS) and $M+1$ small cell data base stations (DBSs). Under this architecture, each UE can connect with CBS and DBS at the same time for control signaling and data transmission, respectively. The set of DBSs $\mathcal{M}$ include one high power node (HPN), providing seamless data transmission coverage with the index of $M+1$, which cover the same area as those $M$ low power nodes (LPNs), supporting high data rate transmission. The set of user equipment (UEs), denoted by $\mathcal{K}=\{1, 2, ..., $\,K$\}$, are served by network. All baseband signals are processed in cloud BBU pool and transmitted to the CBS, and the CBS delivers the baseband signals to each DBS via heterogeneous fronthaul, wired or wireless links. All UE information and channel information needed for resource allocation can be obtained. The carrier frequency of CBS for control signaling is orthogonal with DBSs' for data transmission. The indicator of fronthaul form of DBS was defined as $\bf{w}=$\lbrack$w_1$,$\,w_2$,\,...,$\,w_{M+1}$\rbrack. Here $w_m$ can only be 1 or 0, indicates whether or not the DBS is equipped with wireless fronthaul. We assume all DBSs have single antenna. The total bandwidth $B$ for data transmission is shared among DBSs through orthogonal frequency duplex multiple access (OFDMA), which is divided into a set of $\mathcal{N}=\{1, 2, ... , $\,N$\}$ resource blocks (RBs). We consider distance-dependent path loss, shadowing effect, and the small-scale fading. The small-scale fading is assumed to be frequency-selective. Each RB can only be exclusively assigned to one UE at most in each resource allocation interval.
\subsubsection{Achievable Sum-Rate}
Denote $p_{k,m,n}$ and $g_{k,m,n}$ as the allocated power and channel gain for BS $m$ to UE $k$ on the $n$th RB, respectively. Then the maximum achievable data rate ${R}_{k,m,n}$ of BS $m$ to UE $k$ on the $n$th RB can be calculated as
\begin{equation}
{R}_{k,m,n}  = B_0 \log _2 \left( {1 + \frac{{p_{k,m,n} g_{k,m,n} }}{{B_0 N_0 }}} \right),
\label{achieve_rate}
\end{equation}
where $B_0=B/N$ denotes the bandwidth of each RB, $N_0$ is the spectral density of additive white Gaussian noise (AWGN). In this paper, due to the overlay mechanism, the interference between HPN and LPNs can be totally eliminated. Also we assume the interference from neighbor LPNs can be treated as noise for the sake of simplicity.

We define a binary variable  as RB assignment indicator as
\begin{equation}
\alpha _{k,m,n}  = \left\{ {\begin{array}{*{20}c}
   {1,} \hfill & {n{\rm{th\,RB\,of\,\it{m}\rm{th}\,BS}} {\rm{\,for\, \it{k}}\rm{th\,UE}} ,} \hfill  \\
   {0,} \hfill & {{\rm{otherwise}}{\rm{.}}} \hfill  \\
\end{array}} . \right.
\label{assign_indicator}
\end{equation}

Then, we construct two $K \times(M+1)\times N$ matrix variables, ${\bf{A}}=[\alpha _{k,m,n}]_{K \times (M + 1) \times N}$ and ${\bf{P}}=[p_{k,m,n} ]_{K \times (M + 1) \times N}$, as the feasible RB assignment and power allocation for DBSs, respectively. Therefore, the total sum-rate of the system is expressed as
\begin{equation}
{R}_{\rm{T}} \left( {{\bf{A}},{\bf{P}}} \right){\rm{ = }}\sum\limits_{k = 1}^K {\sum\limits_{m = 1}^{M + 1} {\sum\limits_{n = 1}^N {\alpha _{k,m,n} {R}_{k,m,n} } } } .
\label{total_rate}
\end{equation}
\subsubsection{Fronthaul Model}
In H-CRAN model, each DBS connects with CBS via heterogeneous fronthaul, wired or wireless links. In this paper, the wired fronthaul is assumed to  be ideal medium, has sufficient capacity to deliver data from CBS to DBSs. Since the wireless fronthaul links is out-of-band, there is no interference between the wireless fronthaul links and wireless access links. The focus of our discussion is that, even the wireless fronthaul links can provide huge capacity, mainly benefits from the development of mmWave technique \cite{ref7}, the overall capacity of wireless fronthaul links is still restrained due to the limitation of spectral bandwidth.
\subsection{Optimization Problem Formulation}
\subsubsection{Power Consumption Model}
In this section, we try to figure out how much power consumption can be saved in the CDSA networks compared to the conventional networks. 

Firstly, the power consumption model of H-CRAN is totally different from the conventional networks. In H-CRAN, the baseband units and fronthaul occupy large proportion of system power consumption. As \cite{ref17} presents, the total power consumption  of H-CRAN is given by
\begin{equation}
P_{\rm{T}}  = P_{\rm{S}}  + P_{\rm{F}}  + \sum\limits_{i \in {\cal M}}^{} {P_{\rm{BS}}^{(i)} } ,
\label{total_power}
\end{equation}
where $P_{\rm{S}},\,P_{\rm{F}}$, and $P_{BS}^{(i)}$ denote the power consumption in cloud-based baseband platform, fronthaul links, and BSs, respectively. In a simple classification, all of power consumption can divided into static power and dynamic power, which depends on system load. 

Therefore, to simplify calculation, it is reasonable to approximate the complicate formula into relatively concise and generalized form as 
\begin{equation}
{P}_{\rm{T}}  = \phi _{\rm{E}} \sum\limits_{k = 1}^K {\sum\limits_{m = 1}^{M + 1} {\sum\limits_{n = 1}^N {\alpha _{k,m,n} p_{k,m,n} } } }  + P_{\rm{Static}} ,
\label{transformed_power}
\end{equation}
where $\phi_{\rm{E}}$ denotes the efficiency factor related to network components like power amplifier, fronthaul technique and central processing units, and $P_{\rm{Static}}$ denotes the static power consumption of overall system, including static power consumption of fronthaul, BSs and signal processing platform.

Secondly, compared to convention architecture, the reduction of power consumption of the CDSA networks can be summarized into three points. First of all, due to the centralization of control signaling, the power consumption of control signals in DBSs over the air can be reduced significantly \cite{ref16}, without notable increment of control signals burden in CBS. Then, the baseband component in conventional BS will  disappear in H-CRAN architecture since the processing function of BS is transferred to cloud BBU pool. Moreover, because of great reduction of  control signals between the CBS and DBSs, the overhead traffic of fronthaul will decrease approximately  $10\%$ according to \cite{ref18}. For the power consumption of signal processing, it is the same as that of the conventional networks. Other specific improvement on DBS's power consumption can be fulfilled, because DBS doesn't need always-on and transmit control signals. This parts of power reduction will not be discussed in this paper.
\subsubsection{Problem Formulation}
Our target is to find the optimal ${\bf{A}}$ and ${\bf{P}}$ to maximize EE of overall system, while satisfying the constraints of average minimum data rate of each UE, maximum wireless fronthaul capacity and maximum transmission power of each DBS. Specifically, the objective function is given by
\begin{equation}
\eta _{\rm{EE}} \left( {{\bf{A}},{\bf{P}}} \right) = \frac{{{ R}_{{\rm{T}}} \left( {{\bf{A}},{\bf{P}}} \right)}}{{{ P}_{{\rm{T}}} \left( {{\bf{A}},{\bf{P}}} \right)}}
\label{energy_efficiency}.
\end{equation}

Therefore, the optimization problem can be mathematically formulated as 
\begin{equation}
\begin{array}{*{20}c}
   {\mathop {\max }\limits_{\{ {\bf{A}},{\bf{P}}\} } } & {\eta _{\rm{EE}} }  \\
   {\rm{s.t.}} & {}  \\
   {C_1:} & {\sum\limits_{m = 1}^{M + 1} {\sum\limits_{n = 1}^N {\alpha _{k,m,n} { R}_{k,m,n} } }  \ge r_k^{\min } {\rm{,}}\forall k}  \\
   {C_2:} & {\sum\limits_{k = 1}^K {\sum\limits_{N = 1}^N {\alpha _{k,m,n} p_{k,m,n} } }  \le P_m^{\max } ,\forall m}  \\
   {C_3:} & {\sum\limits_{m = 1}^{M + 1} {\omega _m \sum\limits_{k = 1}^K {\sum\limits_{n = 1}^N {\alpha _{k,m,n} { R}_{k,m,n} } } }  \le R_{\rm{F}}^{\max } }  \\
   {C_4:} & {\sum\limits_{k = 1}^N {\alpha _{k,m,n} }  \le 1,\forall m,n}  \\
   {C_5:} & {\alpha _{k,m,n}  = \{ 0,1\} ,\forall k,m,n}  \\
\end{array}
\label{optimization_problem}
\end{equation}
where $C_1$ is the constraints of average minimum data rate requirement of each UE $r_k^{\min }$, $C_2$ denotes the constraints of transmit power of each BS $P_m^{\max }$, $C_3$ is the constraint of maximum wireless fronthaul capacity limit $ R_{\rm{F}}^{\max }$, $C_4$ and $C_5$ are the intrinsic limitation of RB allocation indicator. Intuitively, the optimization problem is obviously non-convex, and it is difficult to be solved directly with classical convex optimization methods and tools.
\section{Energy Efficient Resource Allocation Algorithm}
\subsection{Optimization Problem Transformation}
We notice that the optimization problem (\ref{optimization_problem}) can be classified as nonlinear fractional programming problem. So we can transform the problem using the well-known Dinkelbach method \cite{ref19}. For the sake of notational simplicity, we defined the optimal EE $q^*$ of the optimization problem as
\begin{equation}
q^*  = \frac{{{ R}_{{\rm{T}}} \left( {{\bf{A}}^* ,{\bf{P}}^* } \right)}}{{{ P}_{{\rm{T}}} \left( {{\bf{A}}^* ,{\bf{P}}^* } \right)}} = \mathop {\max }\limits_{\{ {\bf{A}},{\bf{P}}\} } {\rm{ }}\frac{{{ R}_{{\rm{T}}} \left( {{\bf{A}},{\bf{P}}} \right)}}{{{ P}_{{\rm{T}}} \left( {{\bf{A}},{\bf{P}}} \right)}}
\label{fraction_transformation_1}
\end{equation}

\newtheorem{theorem}{\bf{Theorem}}
\begin{theorem}
 the optimal EE $q^*$ is achieved if and only if\\
\begin{equation}
\begin{split}
\mathop {\max }\limits_{\{ {\bf{A}},{\bf{P}}\} } \,\,\,&{ R}_{{\rm{T}}} \left( {{\bf{A}},{\bf{P}}} \right) - q^* { P}_{{\rm{T}}} \left( {{\bf{A}},{\bf{P}}} \right) \\
= \,\,\,&{ R}_{{\rm{T}}} \left( {{\bf{A}}^* ,{\bf{P}}^* } \right) - q^* { P}_{{\rm{T}}} \left( {{\bf{A}}^* ,{\bf{P}}^* } \right) = 0,
\end{split}
\label{fraction_transformation_2}
\end{equation}
for\,${ R}_{{\rm{T}}} \left( {{\bf{A}},{\bf{P}}} \right) \ge 0$ and ${ P}_{{\rm{T}}} \left( {{\bf{A}},{\bf{P}}} \right) \ge 0$, where $\{ {\bf{A}},{\bf{P}}\}$ is any feasible solution to satisfy the constraints of optimization problem.
\end{theorem}
\begin{IEEEproof}
the proof has been presented in \cite{ref20}. For brevity, the proof is omitted here. 
\end{IEEEproof}

Based on $\bf {Theorem\,1}$, the problem (\ref{optimization_problem}) is transformed as 
\begin{equation}
\begin{array}{*{20}c}
   {\mathop {\max }\limits_{\{ {\bf{A}},{\bf{P}}\} } } & {{ R}_{\rm{T}} \left( {{\bf{A}},{\bf{P}}} \right) - q \cdot { P}_{\rm{T}} \left( {{\bf{A}},{\bf{P}}} \right)}  \\
   {s.t.} & {C_1 \sim C_4,{\rm{ }}\alpha _{k,m,n}  = [0,1],\forall k,m,n}.  
\end{array}
\label{transformed_problem}
\end{equation}
where we change the domain of variables $\alpha _{k,m,n}$ into continuous domain $[0,1]$. The solution to the relaxed optimization problem is actually an upper bound.

\subsection{Energy Efficient Resource Allocation Algorithm}
The optimization problem (\ref{transformed_problem}) is still non-convex since the existence of $C_3$. In general, if the Lagrangian dual decomposition method is used to solve the problem (\ref{transformed_problem}), the duality gap is not zero because of its non-convexity. However, it has been proved that if the problem satisfies some specific conditions, the so-called time-sharing conditions \cite{ref21}, the duality gap is nearly zero. Therefore, we can solve the optimization problem with Lagrangian dual decomposition method.

We construct the Lagrangian function of problem as
\begin{equation}
\begin{array}{l}
 L\left( {{\bf{A}},{\bf{P}},{\bm{\mu}},{\bm{\gamma}} ,{{\upsilon}} } \right) = \sum\limits_{k = 1}^K {\sum\limits_{m = 1}^{M + 1} {\sum\limits_{n = 1}^N {\alpha _{k,m,n} { R}_{k,m,n} } } }  \\ 
  - q\left( \phi _{\rm{E}} {\sum\limits_{k = 1}^K {\sum\limits_{m = 1}^{M + 1} {\sum\limits_{n = 1}^N {\alpha _{k,m,n}  } p_{k,m,n} } }  + P_{\rm{Static}} } \right) \\ 
  + \sum\limits_{k = 1}^K {\mu _k \left( {\sum\limits_{m = 1}^{M + 1} {\sum\limits_{n = 1}^N {\alpha _{k,m,n} { R}_{k,m,n} } }  - r_k^{\min } } \right)}  \\
  - \sum\limits_{m = 1}^{M + 1} {\gamma _m \left( {\sum\limits_{k = 1}^K {\sum\limits_{n = 1}^N {\alpha _{k,m,n} p_{k,m,n} } }  - P_m^{\max } } \right)}  \\  
  - \upsilon \left( {\sum\limits_{m = 1}^{M{\rm{ + }}1} {\omega _m \sum\limits_{k = 1}^K {\sum\limits_{n = 1}^N {\alpha _{k,m,n} { R}_{k,m,n} } } }  - R_{\rm{F}}^{\max } } \right) \\ 
  {\rm{s.t.}}{\rm{\,\,\,\,\,\,\,\,\,\,\,\,C_4,\,}}\alpha _{k,m,n}  = [0,1],\forall k,m,n. \\
 \end{array}
\label{Lagraian_function}
\end{equation}
where $\bm{\mu}$ is the Lagrange multiplier vector of dimension $K$ corresponding to the average minimum data rate of UEs, and $\bm{\gamma}$ is the Lagrange multiplier vector of dimension $M+1$ corresponding to the maximum transmission power of BS. The variable ${\upsilon}$ is the Lagrange multiplier corresponding to maximum wireless fronthaul capacity. Besides, the boundary constraints respect to $\alpha _{k,m,n}$ will be absorbed into the optimal solution in the following procedure.

The Lagrangian dual function is 
\begin{equation}
\begin{array}{*{20}c}
   {g(\bm{\mu},\bm{\gamma} ,{\upsilon})} \hfill & { = \mathop {\max }\limits_{{\rm{\{ }}{\bf{A}},{\bf{P}}{\rm{\} }}} {{L}}({\bf{A}},{\bf{P}},\bm{\mu},\bm{\gamma} ,{\upsilon} )} \hfill  \\
   {s.t.} \hfill & {C_4,\alpha _{k,m,n}  = [0,1],\forall k,m,n}, \hfill  \\
\end{array}
\label{Dual_function}
\end{equation}
and the dual optimization problem is
\begin{equation}
\begin{array}{*{20}c}
   {\mathop {\min }\limits_{\{ \bm{\mu},\bm{\gamma} ,{\upsilon} \} } } & {g\left( {\bm{\mu},\bm{\gamma} ,{\upsilon}} \right)}  \\
   {s.t.} & {\bm{\mu}  \ge 0,\bm{\gamma}  \ge 0,{\upsilon}  \ge 0}.  \\
\end{array}
\label{Dual_optimization_problem}
\end{equation}

Note that the dual optimization problem is always convex, so the dual decomposition method can be used to solve this dual problem. We decompose the problem into $N$ independent sub-problems with identical structure as
\begin{equation}
\begin{split}
g \left( {\bm{\mu},\bm{\gamma} ,{\upsilon}} \right)&= \mathop {\max }\limits_{{\rm{\{ }}{\bf{A}},{\bf{P}}{\rm{\} }}} \left\{ {\sum\limits_{n = 1}^N {g_n \left( {\bm{\mu},\bm{\gamma} ,{\upsilon}} \right)}  - q\cdot P_{\rm{Static}} } \right.\\
&\left. { - \sum\limits_{k = 1}^K {\mu _k \cdot r_k^{\min } }  + \sum\limits_{m = 1}^{M + 1} {\gamma _m \cdot P_m^{\max } }  + \upsilon \cdot R_{\max } } \right\},
 \end{split}
\label{sum_problem}
\end{equation}
where
\begin{equation}
\begin{split}
&g_n \left( {\bm{\mu},\bm{\gamma} ,{\upsilon}} \right) =\\
&\mathop {\max }\limits_{\{ {\bf{A,P}}\} } \sum\limits_{k = 1}^K {\sum\limits_{m = 1}^{M + 1} {\left[ \begin{array}{l}
(1 + {\mu _k} - \upsilon  \cdot {\omega _m}) \cdot {\alpha _{k,m,n}} \cdot {R_{k,m,n}} - \\
({\gamma _m} + q \cdot {\phi _E}) \cdot {\alpha _{k,m,n}} \cdot {p_{k,m,n}}
\end{array} \right]} }.
 \end{split}
\label{sub_problem}
\end{equation}

Note that $g_n \left( {\bm{\mu},\bm{\gamma} ,{\upsilon} } \right)$ is convex with respect to $p_{k,m,n}$, with the Karush-Kuhn-Tucker (KKT) condition, the optimal power allocation can be calculated as
\begin{equation}
p_{k,m,n}^*  = \left[ {\frac{{B_0 }}{{\ln 2}} \cdot \frac{{1 + \mu _k  - \upsilon \cdot \omega _m }}{{\gamma _m  + q\cdot\phi _{\rm{E}} }} - \frac{{B_0 N_0 }}{{g_{k,m,n} }},0} \right]^ +  ,
\label{optimal_power}
\end{equation}
where $\left[ x \right]^+$ denotes $\max \{ x,0\}$. The optimal power allocation (\ref{optimal_power}) is in the form of multi-level water filling. The water-level depends on both dual variables and individual weight value. Specifically, $\upsilon\cdot\omega _m $ controls the power allocation only for the wireless fronthaul DBSs so that the total capacity of wireless fronthaul will not exceed the limitation; $\mu_k$ controls the average minimum data rate requirement of each UE, the higher requirement will result in higher water-level.

Then, substituting the optimal power allocation obtained in (\ref{optimal_power}) into (\ref{sub_problem}),  the optimal RB assignment can be decided by
\begin{equation}
\alpha _{k,m,n}^*  = \left\{ {\begin{array}{*{20}c}
   {1,} \hfill & {k= \arg \mathop {\max }\limits_{1 \le k \le N}{\rm{ }}H_{k,m,n} \,{\rm{and}}\,H_{k,m,n}>0,}\hfill\\
   {0,} \hfill & {\rm{otherwise,}} \hfill  \\
\end{array}} ,\right.
\label{optimal_assignment}
\end{equation}
where
\begin{equation}
\begin{split}
H_{k,m,n}  &= \left( {1 + \mu _k  - \upsilon \omega _m } \right){ R}_{k,m,n} (p_{k,m,n}^* )\\
&- \left( {\gamma _m  + q\cdot \phi _{\rm{E}} } \right) p_{k,m,n}^*
 \end{split}.
\label{H}
\end{equation}

After we obtain the optimal power allocation and RB assignment, we use the sub-gradient method \cite{ref22} to solve the dual problem.
The sub-gradient of dual function is given by
\begin{align}
&\nabla \mu _k (n) = \sum\limits_{m = 1}^{M + 1} {\sum\limits_{n = 1}^N {\alpha _{k,m,n} { R}_{k,m,n} } }  - r_k^{\min } ,\forall k,\\
&\nabla \gamma _m (n) = P_m^{\max } {\rm{ - }}\sum\limits_{k = 1}^K {\sum\limits_{n = 1}^N {\alpha _{k,m,n} p_{k,m,n} } } ,\forall m,\\
&\nabla \upsilon (n) = R_{\max }  - \sum\limits_{m = 1}^{M{\rm{ + }}1} {\omega _m \sum\limits_{k = 1}^K {\sum\limits_{n = 1}^N {\alpha _{k,m,n} { R}_{k,m,n} } } } ,
\end{align}

The dual variables in the $n$th iteration are updated by
\begin{align}
&\mu _k (n) = \left[ {\mu _k (n-1) - \delta _\mu  (n) \cdot  \nabla \mu _k (n)} \right]^ +  ,\forall k,\\
&\gamma _m (n) = \left[ {\gamma _m (n-1) - \delta _\gamma  (n)\cdot  \nabla \gamma _m (n)} \right]^ +  ,\forall m,\\
&\upsilon (n) = \left[ {\upsilon (n-1) - \delta _\upsilon  (n)\cdot  \nabla \upsilon (n)} \right]^ +  ,
\end{align}
where $\delta _\mu  (n),~\delta _\gamma  (n)$, and $\delta _\upsilon  (n)$ are the positive appropriate small step sizes, respectively.


\section{Simulation Results and Analysis}
In this section, the EE performance of proposed energy efficient resource allocation algorithm is evaluated with simulation. We consider the CDSA networks in a hexagonal deployment with the inter-site distance (ISD) of 500 meters. CBS and HPN located in the central for simplicity, within which LPNs and UEs are randomly and uniformly distributed. The carrier center frequency of HPN and LPNs is 2 GHz and 3.5 GHz. The total system bandwidth is 10 MHz divided into 50 RBs. Without specialization, the default value of the number of wireless fronthaul DBSs, total DBSs and UEs is 10, 20 and 50, respectively. And the average minimum data rate requirement of UEs is 5 Mbps, the overall wireless fronthaul capacity is 0.8 Gbps (medium constraint). It is assumed that the path loss model is $128.1 + 37.6\log _{10} (d)$ for links between HPN and UEs, and $140.7 + 36.7\log _{10} (d)$ for the links between LPNs and UEs, where $d$ denotes the distance between the DBS and UE in {\it{km}}. The shadowing deviation of HPN and LPNs is 8 dB and 10 dB. The noise power spectral density $N_0$ is -174 dBm/Hz.
 
We adopt the power consumption model and parameters of DBS as the same as that of \cite{ref23}. The maximum transmit power of HPN and LPNs is 20 W and 0.13 W. The power consumption model of H-CRAN is the same as that of \cite{ref17}. In the modified power consumption model, the value of $\phi _{\rm{E}}$ is 0.29, and the static power consumption $P_{\rm{Static}}$ is 439 W. The overhead cost of DBSs is  $28\%$ according to \cite{ref16}. We assume that the power consumption of fronthaul in the CDSA networks is $10\%$ less than the conventional networks, and that both the overhead cost and baseband consumption of DBSs are zero, as described in the Section II.
 \begin{figure}[!t]
\centering
 \includegraphics[width=3.5in]{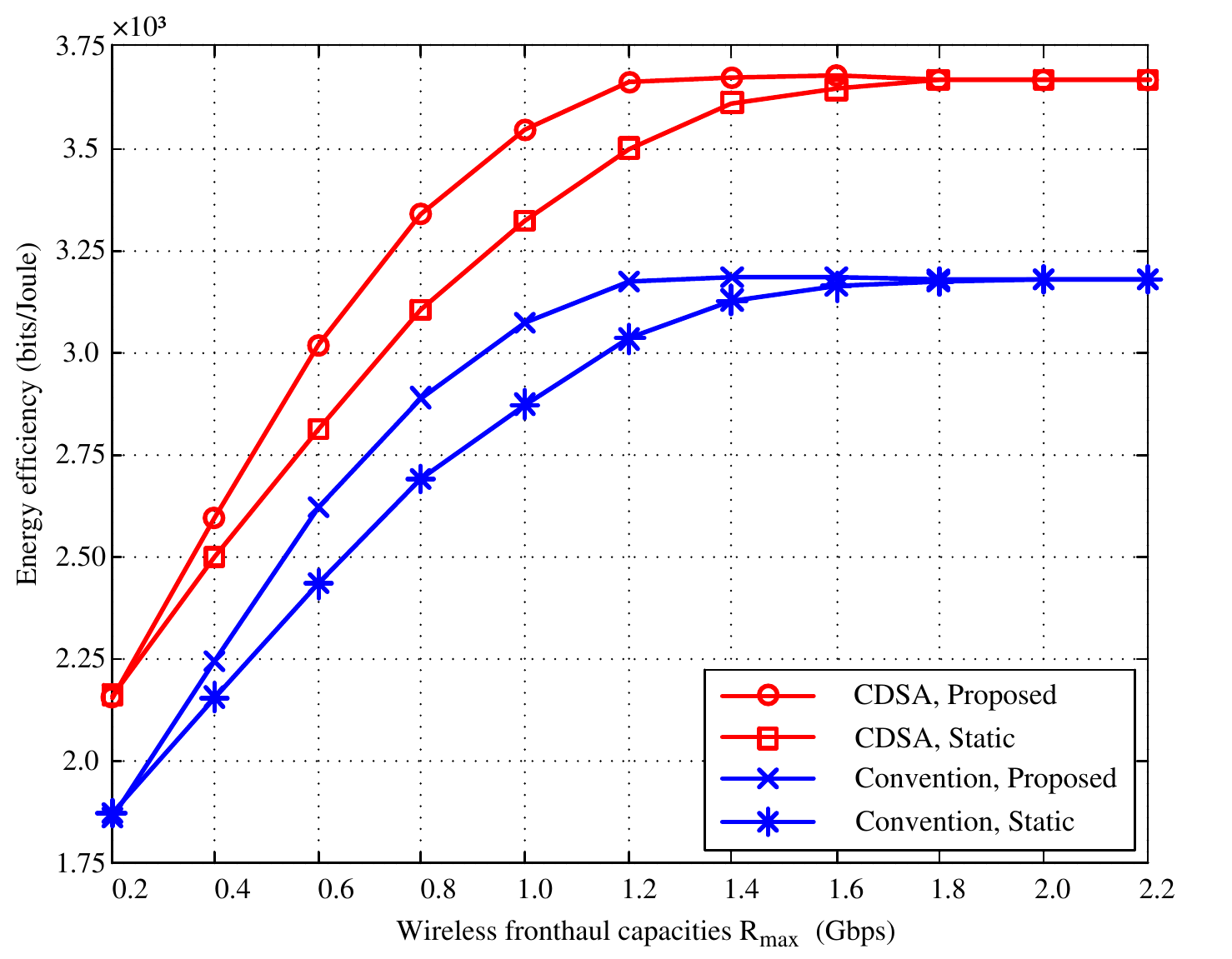}
     \caption{Energy efficiency vs.  wireless fronthaul capacity}
     \label{FronthaulCapacity}
\end{figure}

Figure 1 demonstrates the impact of wireless fronthaul capacity on network EE with different network architectures. The static algorithm means that all wireless fronthaul DBSs are allocated with same fronthaul capacity. It is observed that, with the increment of wireless fronthaul capacity, the network EE increases gradually due to the loosen capacity limitation. Meanwhile, benefitting from the dynamic allocation algorithm of fronthaul capacity, the gap of EE gain between proposed algorithm and static algorithm becomes larger. Also, for the sufficient wireless fronthaul capacity, e.g., 2 Gbps, different algorithm achieves the same and maximum EE. But our proposed algorithm achieves the maximum EE with less wireless fronthaul capacity than the static algorithm, because of its efficiency of fronthaul capacity allocation. Furthermore, the CDSA networks have significant improvement in terms of network EE compared with the conventional network.


In Fig. 2, the difference of network EE of various algorithms and network architectures are presented,  with gradually increasing numbers of wireless fronthaul DBSs. In the static algorithm, maximum data rate of each wireless fronthaul DBS is assumed to be 50 Mbps. In proposed algorithm, the overall wireless fronthaul capacity is allocated dynamically. It can be seen that the network EE decreases with the increasing the number of wireless fronthaul DBSs. Furthermore, the more wireless fronthaul DBSs, the larger gap between our proposed algorithm and static algorithm, which comes from the dynamic allocation scheme of fronthaul capacity. 
\begin{figure}[!t]
\centering
     \includegraphics[width=3.5in]{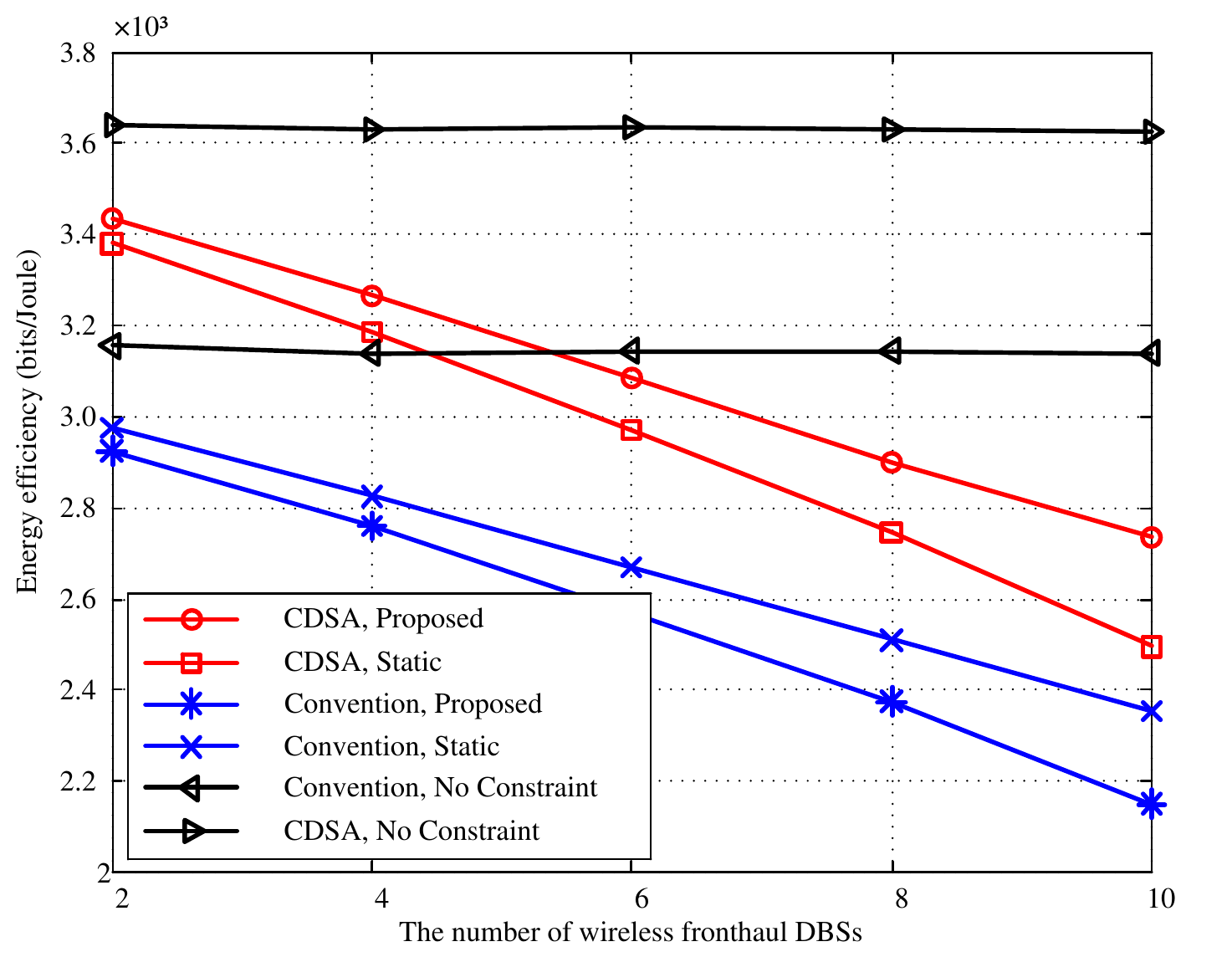}
     \caption{Energy efficiency vs. the number of wireless fronthaul DBSs}
     \label{FronthaulBSNumber}
\end{figure}

In Fig. 3, the impact of average minimum data rate requirement of UEs on network EE is shown. Because the system has to perform inefficient allocation in order to satisfy the constraints of UEs, along with the increment of average minimum data rate requirement of UEs, the network EE decrease gradually.  The lower UEs' data rate require, the less constraints the system has, and the more degree of freedom can be utilized to enhance the network EE. As expected, the network EE of proposed algorithm is higher than that of static algorithm, and the CDSA networks achieve higher EE compared to the conventional networks. In this figure, we do not show the results when the average minimum data rate requirement is bigger than 8 Mbps since the demand of UEs are beyond the  system capacity. 

In Fig. 4, the relationship between the deployment density of DBS and the SE/EE for different networks are shown for 50 UEs within whole coverage area.Because the difference of networks is power consumption model, the SE of different networks is the same. When density is lower than 200 per km$^2$, the SE increases along with the rising of DBS density significantly, and then the SE almost keeps the same. Because there is a minimum distance between UE and its associated DBS predefined in the simulation settings, when DBS number increases above one threshold, the received SNR will no longer increase. So that the SE cannot be improved significantly. On the other hand, for the CDSA networks, the network EE rises rapidly and then declines slowly because of the increased power consumption of DBSs. Besides, It can be observed that the EE of CDSA networks is greater than that of conventional networks due to the saved power consumption.

\section{Conclusion}
In this paper, we investigated the impact of heterogeneous fronthaul and control data separation architecture (CDSA) on the  network EE in downlink of H-CRAN. Aiming to maximum the network EE, we formulated the optimization problem with modified power consumption model. By fractional programming method and Lagrangian dual decomposition method, we derived the close-form optimal solution. Simulation results show that our proposed algorithm has maximum $8\%$ EE gain  compared to conventional static algorithm, and the CDSA  can produce maximum $16\%$ EE gain compared to the conventional network architecture.

\begin{figure}[!t]
\centering
     \includegraphics[width=3.5in]{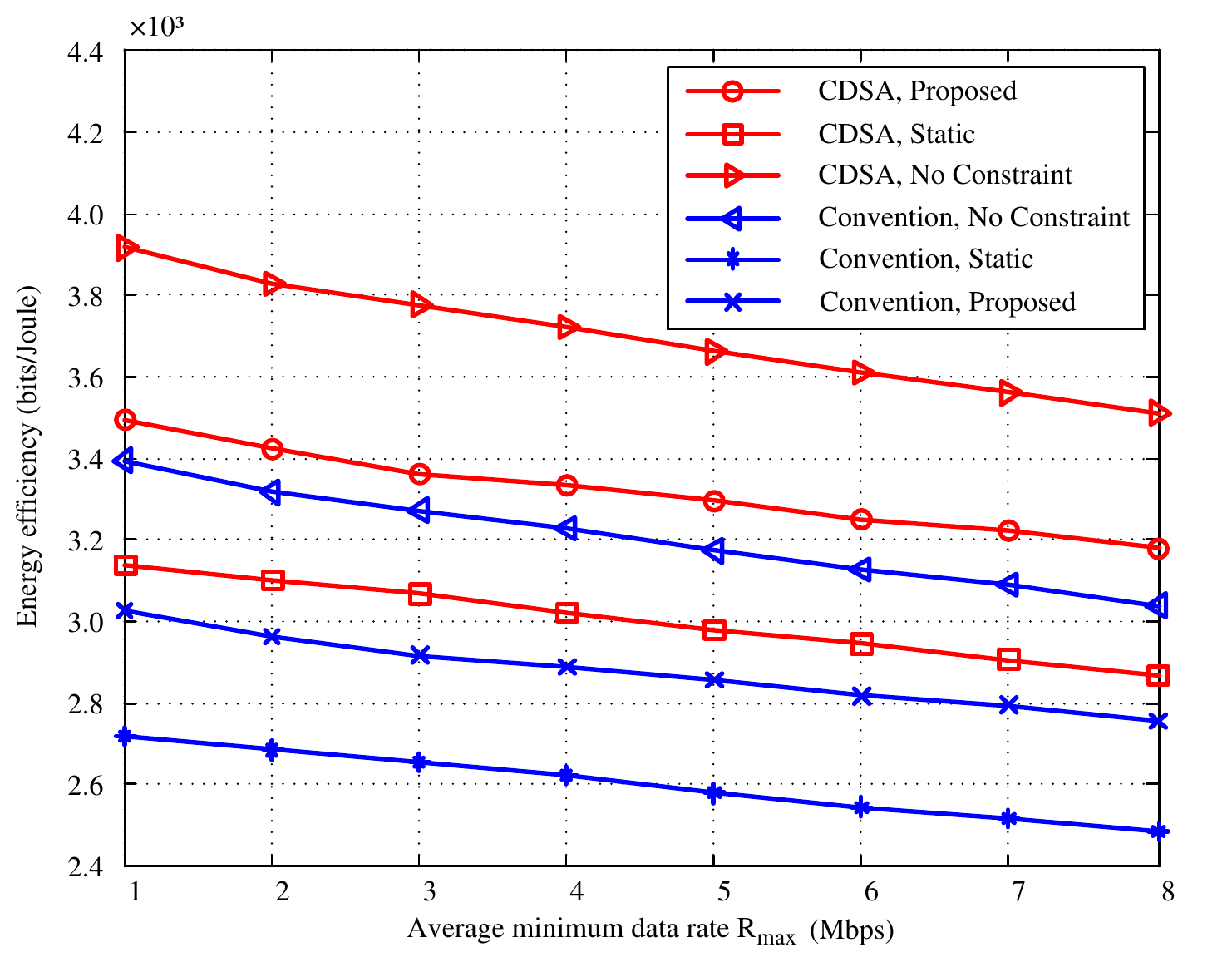}
     \caption{Energy efficiency versus average minimum data rate requirement of UEs}
     \label{MinRate}
\end{figure}

 \begin{figure}[!t]
	\centering
     \includegraphics[width=3.5in]{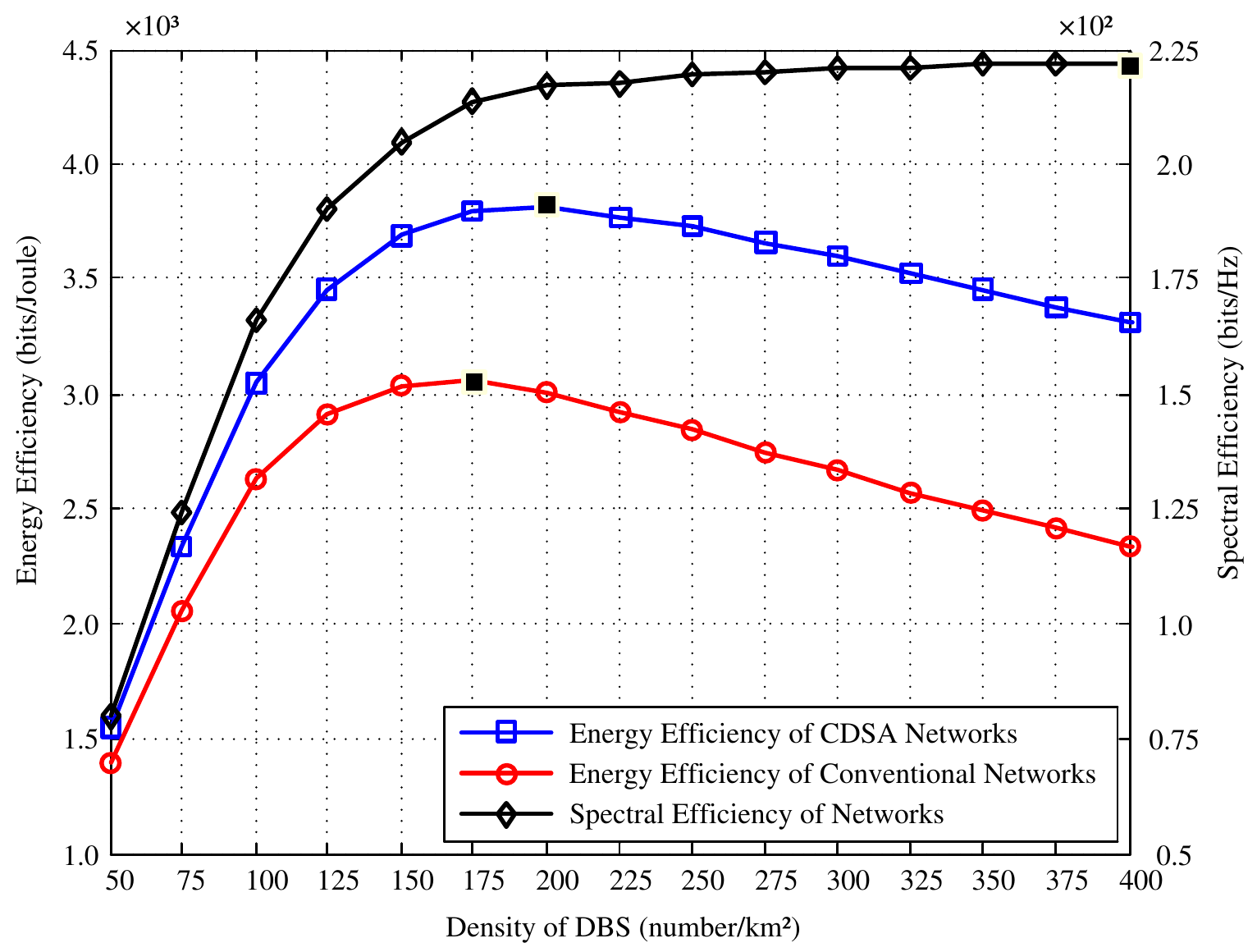}
     \caption{Energy efficiency versus the density of DBS}
     \label{DBSNumber}
 \end{figure}


\section*{Acknowledgment}

The paper is in part supported by the National Basic Research Program of China (973 Program) under grant 2012CB316003.


%

\end{document}